\documentclass{iopart}

\usepackage[utf8]{inputenc}
\usepackage[T1]{fontenc}
\usepackage{graphicx}
\usepackage{placeins}
\usepackage{color}

\graphicspath{{./pics/}}

\begin{document}

\title[Fast time-domain measurements on telecom single photons]{Fast time-domain measurements on telecom single photons}
\author{Markus Allgaier$^1$, Gesche Vigh$^1$, Vahid Ansari$^1$, Christof Eigner$^1$, Viktor Quiring$^1$, Raimund Ricken$^1$, Benjamin Brecht$^{1,2}$ , Christine Silberhorn$^1$}

\newcommand{\correction}[1]{\textcolor{red}{#1}}

\address{$^1$ Integrated Quantum Optics, Applied Physics, University of Paderborn, 33098 Paderborn, Germany}
\address{$^2$ Clarendon Laboratory, Department of Physics, University of Oxford, Oxford OX1 3PU, United Kingdom}
\ead{markus.allgaier@upb.de}
\begin{abstract}
Direct measurements on the temporal envelope of quantum light are a challenging task and not many examples are known since most classical pulse characterisation methods do not work on the single photon level. Knowledge of both spectrum and timing can however give insights on properties that cannot be determined by the spectrum alone. While temporal measurements on single photons on timescales of tens of picoseconds are possible with superconducting photon detectors and picosecond measurements have been performed using streak cameras, there are no commercial single photon sensitive devices with femtosecond resolution available. While time-domain sampling using sum-frequency generation has been already exploited for such measurement, inefficient conversion has necessitated long integration times to build the temporal profile.
We demonstrate a highly efficient waveguided sum-frequency generation process in Lithium Niobate to measure the temporal envelope of single photons with femtosecond resolution with short enough acquisition time to provide a live-view of the measurement.
We demonstrate the measurement technique and combine it with spectral measurements using a dispersive fiber time-of-flight spectrometer to determine upper and lower bounds for the spectral purity of heralded single photons.
The approach complements the joint spectral intensity measurements as a measure on the purity can be given without knowledge of the spectral phase.
\end{abstract}
\maketitle

\section*{Introduction}

One key aspect to assess the quality and usability of single photons is the spectral purity \cite{alan_migdall_single-photon_2013}. The established means of measuring this quantity is through interference with a known reference pulse \cite{cassemiro_accessing_2010}. There is also the possibility to ascertain the photon purity in biphoton states generated through spontaneous parametric down-conversion (PDC) by measuring the joint spectral intensity, but only if no non-linear phases terms on the PDC pump and phasematching function are present \cite{bre2013}. This is due to the fact that one needs the complete spectral or temporal information including phase information, which is hard to measure, even though it is possible in both the spectral and the temporal domain. Complete intensity and phase characterisation of the spectrum of single photons was demonstrated by using spectral shearing interferometry \cite{fittinghoff_measurement_1996,davis_single-photon_2016}. Homodyne detection enables the intensity and phase characterisation of the temporal envelope \cite{qin_complete_2015}.
There is, however, also the option of combining temporal and spectral intensity information. As the time-bandwidth-product (TBP) can be extracted from intensity measurements, it can be determined with relative ease, and it was shown that the time-bandwidth-product contains information on purity \cite{bre2013,uren_generation_2007}. As spectral intensity measurements can be determined both by time-of-flight dispersive fiber spectrometers \cite{avenhaus_fiber-assisted_2009} and commercially available single photon sensitive spectrometers, one only needs a means of temporal intensity characterisation. Among these are homodyne detection \cite{smithey_measurement_1993,anderson_ultrafast_1996}, streak cameras \cite{asmann_ultrafast_2010,wiersig_direct_2009} and ultrafast sampling using sum-frequency generation (SFG) \cite{kuzucu_joint_2008,kuzucu_two-photon_2009}. Temporal measurements by up-conversion sampling are particularly interesting because they can in principle be highly efficient and therefore of measure the temporal envelope with a short integration time. A single photon is up-converted using a sum-frequency generation with a short duration pump pulse. The short pulse has only limited overlap with the single photon and therefore the conversion efficiency depends on the relative timing between the two pulses and, most notably, on the temporal shape of the single photon. As pulse walk-off in non-linear crystals reduces the resolution of the system, existing works have used short non-linear crystals for SFG at the expense of efficiency. Carefully balanced dispersion is necessary to increase the interaction length while maintaining femtosecond resolution

In this work we extend the method of ultrafast up-conversion sampling to long non-linear crystals with dispersion engineering to significantly improve the efficiency of the process and therefore reduce measurement time drastically.
We employ the quantum pulse gate (QPG) \cite{eck11,bre2014a}, a device recently introduced by our group, as a platform to perform these up-conversion measurements. This device is developed around a group-velocity matched, temporal overlap sensitive sum-frequency generation operating with high efficiency \cite{all2016}.
This way we cut down the measurement times from tens of minutes to the order of one second.
In combination with a fast delay device, fast temporal envelope measurements are demonstrated. We combine time-domain measurements with spectral data to determine the purity of single photons generated by a PDC process.

\section*{Methods}

The process we use to sample single photons in the time domain requires a efficient up-conversion that is sensitive to temporal overlap. The conversion process in the QPG employed here takes place in Titanium-indiffused waveguides in Lithium Niobate. In a type-II non-linear process we mix the 1545\,nm single photon input with 854\,nm pump light, generating an output at 550\,nm. The poling period of the 27\,mm long sample is 4.4\,\(\mu\)m. A long interaction length is achieved by matching the group-velocity of the the pump with the group-velocity of the field to be characterised. This is done by counteracting the material and waveguide dispersion with material birefringence. As the two pulses travel at the same speed through the waveguide, the conversion efficiency is highly dependent on the temporal overlap, which stays constant. Therefore, assuming that the pump pulse is sufficiently short in duration, the other input's temporal envelope can be recovered as a function of the delay between the two pulses. This is similar to the method described in \cite{kuzucu_joint_2008}, where a short crystal was used. Conversion efficiency scales with both interaction length and pump pulse energy, and as there are physical limitations for increasing the latter, increasing the interaction length is the only way to increase the efficiency. The group-velocity matching ensures that there is no pulse-walk-off regardless of the crystal length and makes the significantly larger conversion efficiency possible.

\begin{figure}
\centering
\includegraphics[width=0.85\textwidth]{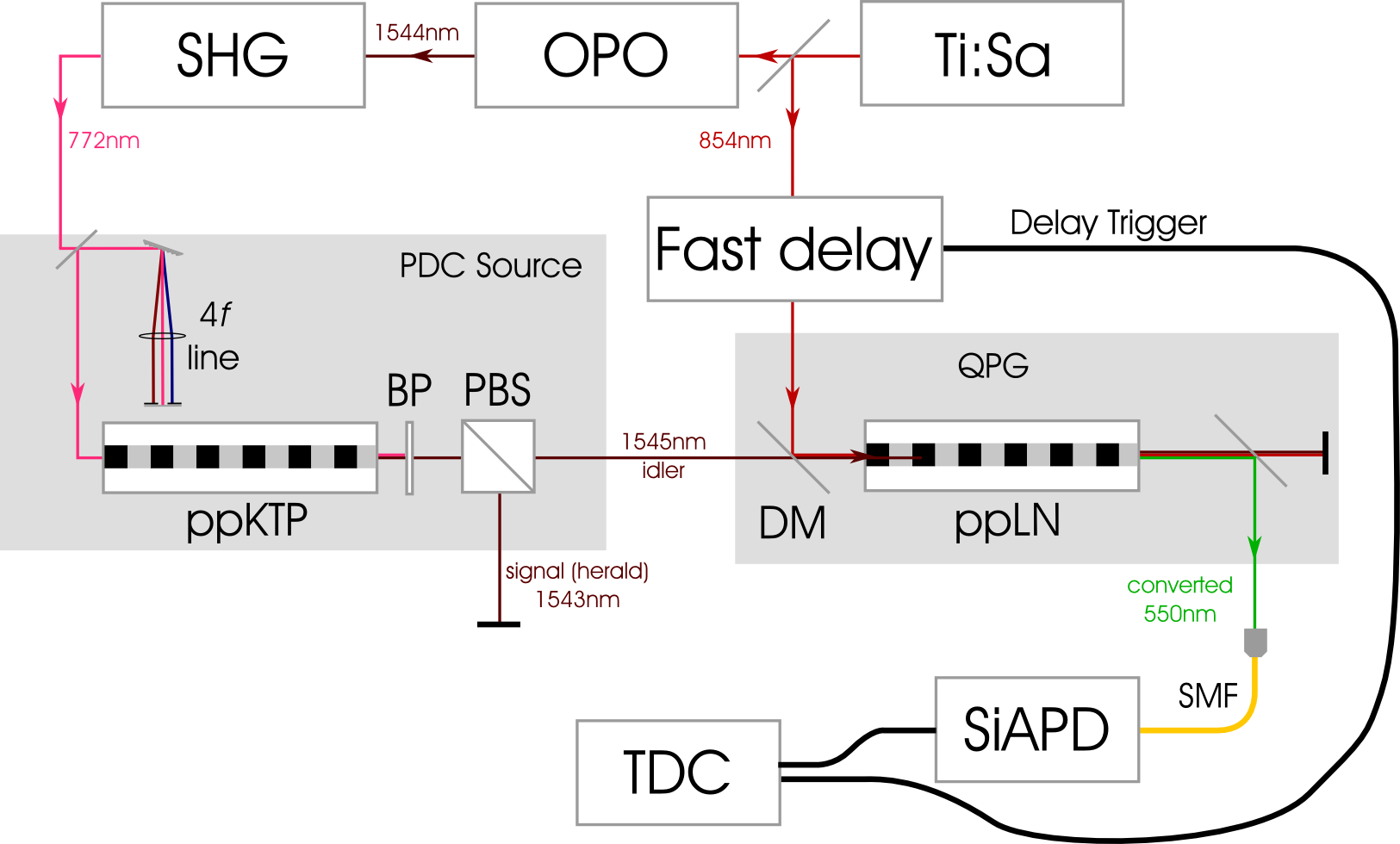}
\caption{Setup used in the experiment. BP: Band pass filter, PBS: Polarizing beam splitter, DM: Dichroic mirror, SMF: Single mode fibre, SiAPD: Silicon Avalanche Photodiode, TDC: Time domain counter}
\label{fig:setup}
\end{figure}

We show our experimental setup in figure~\ref{fig:setup}. We characterise light from a PDC source similar to the one described in Ref. \cite{har2013}. It is a 8mm long KTP crystal with Rubidium exchanged waveguides, poled over a length of 6\,mm with a poling period of 117\,\(\mu\)m. The source is pumped with 772.5\,nm light produced by a cascade of a Coherent Chameleon II Ti:Sapphire laser with a repetition rate of 80.165\,MHz, APE Compact OPO, and a periodically poled bulk Lithium Niobate crystal for second-harmonic generation (SHG). The light has a full-width-at-half-maximum (FWHM) bandwidth of 3\,nm and can be spectrally filtered by means of a folded 4\textit{f} line with a grating and a variable slit in the focal plane of a lens.

The photons from the type-II PDC process are split up by a polarizing beam splitter and coupled into single mode fibers and detected using superconducting nanowire single photon detectors (SNSPDs), achieving a heralding efficiency of 28.1\(\pm\)0.1\,\%. This number is already corrected for the detector efficiency of 90\,\%. The heralded idler photon is then coupled through the QPG where it interacts with the fundamental Ti:Sapphire beam. The light coming out of the QPG is spectrally separated using dichroic mirrors, coupled into fibers and detected using a superconducting nanowire single photon detector (SNSPD) for the unconverted light and a silicon avalanche photodiode (SiAPD) for the converted green light.

\begin{figure}
\centering
\includegraphics[width=0.7\textwidth]{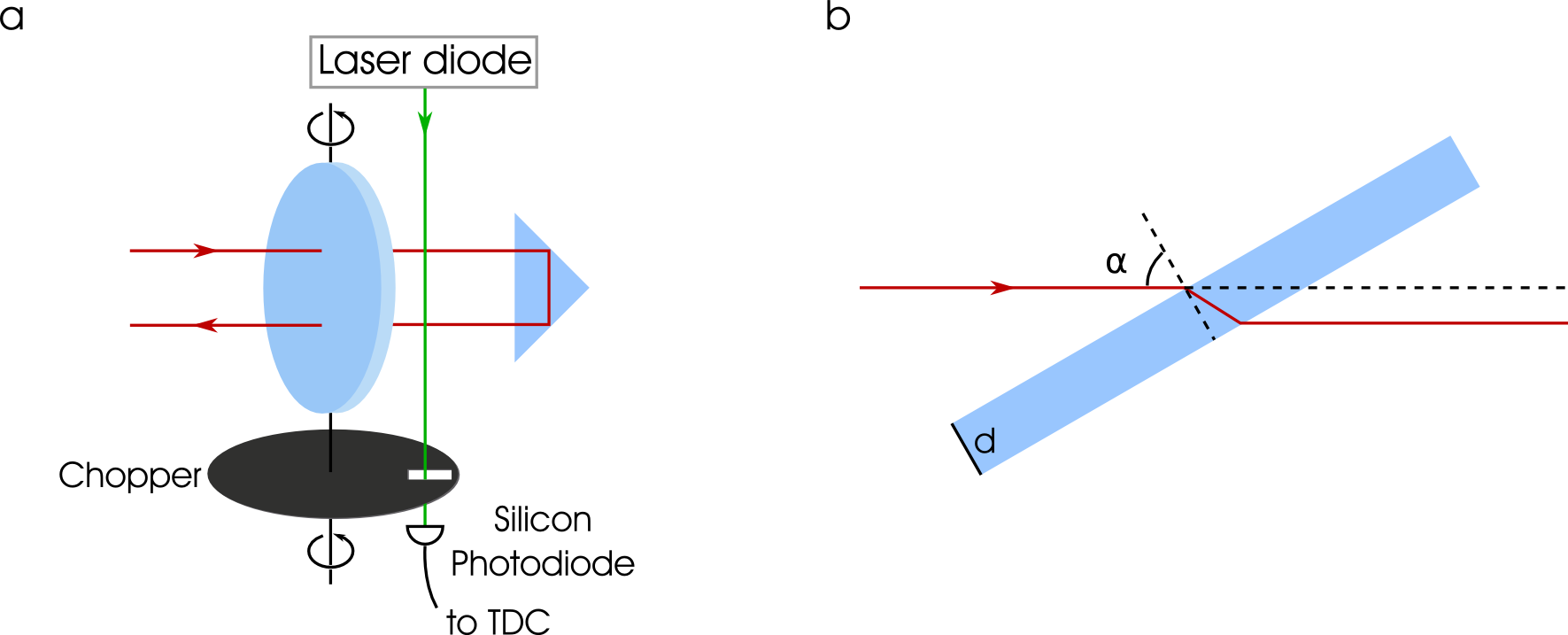}
\caption{(a) Schematic of the fast delay employed in the experiment. The laser beam is coupled through a spinning glass plate making use of the angle dependent optical path. Traversing the plate twice and perpendicular to the rotation axis eliminates beam wandering. The beam from a laser diode, coupled through a chopper wheel and onto a silicon avalanche diode provides a trigger signal for each full rotatation. (b) Optical path through the fast delay device}
\label{fig:fastdelay}
\end{figure}

However, the highly efficient SFG process is only one aspect of why our approach is faster. A second necessity for a quick or even ,,live-view``-like measurement on sub-second timescales is a means of changing the delay between the two fields in a quick, repeatable and controlled way. We identified three ways of managing these requirements: Acousto-optic pulse shapers, fast linear piezo stages, and rotating glass plate delays. Acousto-optic pulse shapers are expensive and have limited wavelength range. Fast linear piezo stages exist, but they are also expensive and the oscillating movement at rates of several Hertz require unreasonable engineering efforts for the optics mounts on top of the stage. The last technique, using a rotating glass plate as a delay, is known from applications such as optical auto-correlation \cite{boggy_rapid_1983} and terahertz imaging \cite{probst_cost-efficient_2014}.  It is depicted in figure~\ref{fig:fastdelay}.

A polished glass plate of 12\,mm thickness is mounted in a 3D-printed frame. The frame, supported by ball bearings, is then driven by a DC motor to rotate at 50\,Hz. The frame also holds a chopper plate that opens the beam pass of a green laser beam only once per rotation. The incident light on a silicon avalanche diode (APD) gives rise to the signal used as a trigger and recorded by the time domain counter (TDC). The pump laser beam for the SFG process passes the plate twice as shown in figure~\ref{fig:fastdelay}b. The double pass. The optical path \(d_{\mathrm{o}}\) through the plate is

\begin{equation}
d_{\mathrm{o}}(\alpha) = \frac{d}{\cos\left( \arcsin \frac{\sin \alpha}{n}\right)}
\end{equation}
where \(\alpha\) is the incidence angle of the beam on the plate and \(n\) is the glass plates refractive index. The plate is made from silica glass with a refractive index of 1.45. When the incidence angle is not zero there is an additional optical path introduced, resulting in the delay
\begin{equation}
\Delta \tau_{\mathrm{max}} =2\cdot  \frac{d_{\mathrm{o}}(\alpha_\mathrm{C})-d}{c_0}\quad .
\end{equation}

%
 \noindent
As there are reflections on the glass surface, the transmitted power depends on the angle. The incident light is in parallel polarisation and we adjust an additional manual delay stage so that the sampling of the waveform takes place when the incidence angle on the glass plate is between perpendicular incidence and the brewster angle. This guarantees a almost constant power level transmitted through the device. Within this bound the delay range of the device is 12.8\,ps which is also the longest waveform that can therefore be sampled by our setup. As the device goes through this delay range four times during one rotation, the waveforms are sampled 200 times per second. From the laser source's repetition rate and the delay device's rotation speed we calculate, that a waveform of this maximum length of 12.8\,ps is sampled with 1000 points, meaning that the distance between two data points in the recovered waveform is 12.8\,fs. The main source of error from this delay device is the fluctuation of the rotation speed. This was assessed by tracking the delay trigger signal frequency from the green laser on a oscilloscope. We then extracted the steepest slope of the curve and calculated to the shift in rotation period. We estimate that the error of the temporal duration of the measured waveform introduced by the delay is at most 70\,fs. The dominating uncertainty in the measurement is the pump pulse duration of 230\,fs. There may be additional contributions such as uneven movement and vibration of the fast delay, or chirp of the sampling pulse. The practical resolution of the method was estimated by performing a measurement on a known reference generated by the optical parametric oscillator. By doing so a measurement uncertainty of 300\,fs was ascertained. As the pump pulse duration as a source of measurement uncertainty is known, we can deconvolve the measurement result and are left with a uncertainty of 200\,fs

The rotational delay trigger signal as well as the single photon detector counts are recorded using a AIT TTM8000 time-to-digital converter (TDC). We recover the single photon waveform in the arrival time histogram of the single photon clicks relative to the rotation trigger signal, where we calculate the corresponding delay to the angle of the delay device.



\section*{Results}
\FloatBarrier

\begin{figure}
\centering
\includegraphics[width=0.45\textwidth]{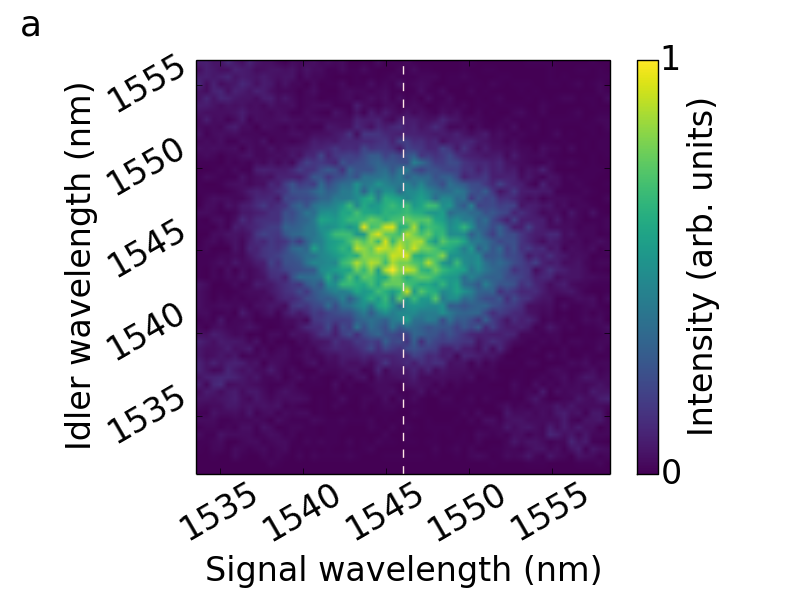}\includegraphics[width=0.45\textwidth]{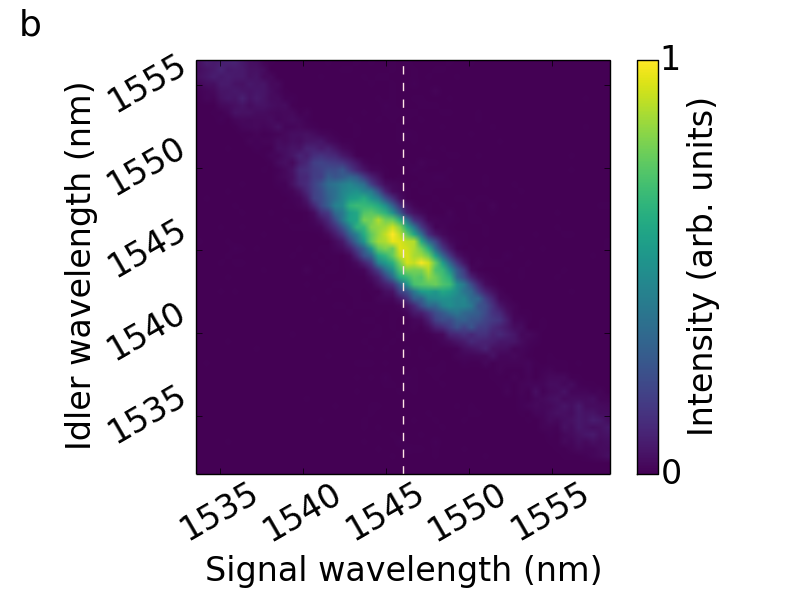}\\
\includegraphics[width=0.45\textwidth]{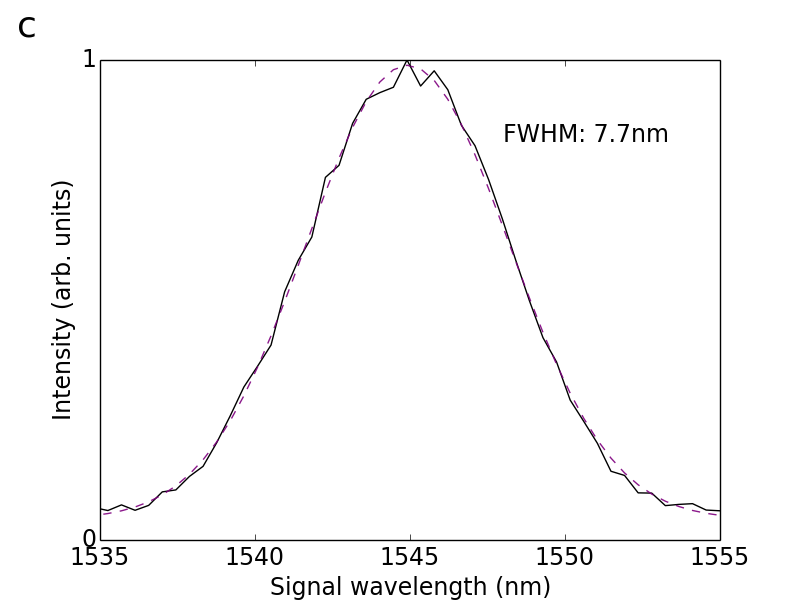}\includegraphics[width=0.45\textwidth]{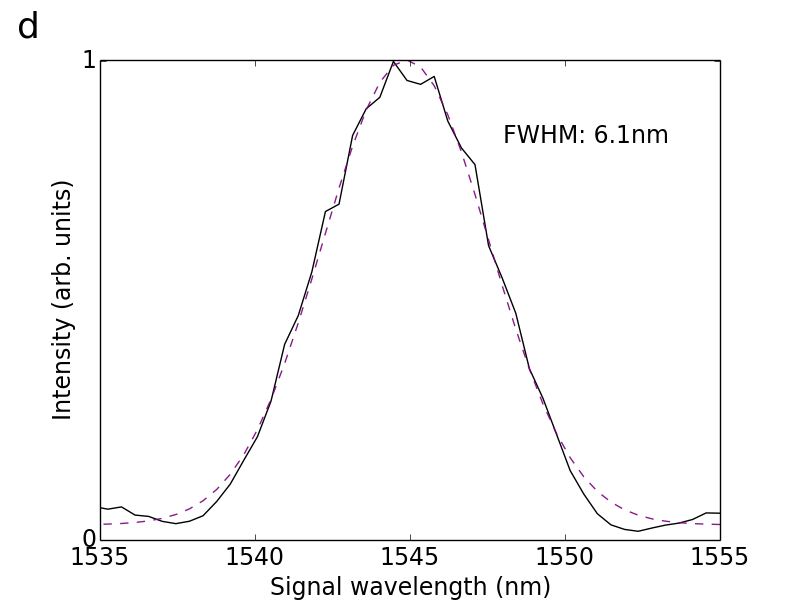}\\
\caption{(a) and (b): Joint spectral intensities of the decorrelated and correlated PDC state measured with a pair of dispersive fiber time-of-flight spectrometers. The dashed lines indicate where the cuts for the calculation of the extected temporal envelopes were taken. (c) and (d); Respective marginal spectral of the idler photons, obteined by integrating the respective JSIs over all signal wavelengths.}
\label{fig:jsi}
\end{figure}

In the experiment we measure two different states from the PDC source. By varying the PDC pump bandwidth we can produce both correlated and decorrelated photon pair states. These are characterised spectrally using a pair of dispersive-fiber time-of-flight spectrometers. The resulting joint spectral intensities are displayed in figure~\ref{fig:jsi}. The decorrelated state is produces with a pump bandwidth of 3.09\,nm, the correlated one with a 0.78\,nm bandwidth pump. Schmidt decomposition of the measured JSIs yields cooperativity numbers of K=1.08 and K=2.10 fo correlated and decorrelated state, respectively. This implies that there is some slight correlation remaining, this is due to the fact that the phasematching angle is not perfectly perpendicular to the pump function in the JSI \cite{eckstein_mastering_2012}. From the JSIs we can extract the spectral bandwidths of the PDC idler photons by extracting the marginal spectra to obtain the FWHM bandwidth. It is \(\Delta\lambda\)=7.7\,nm\(\pm\)0.1\,nm or \(\Delta\nu\)=966\,GHz for the decorrelated PDC state and  \(\Delta\lambda\)=6.1\(\pm\)0.1\,nm or \(\Delta\nu\)=766\,Ghz for the correlated one.

\begin{figure}
\centering
\includegraphics[width=0.45\textwidth]{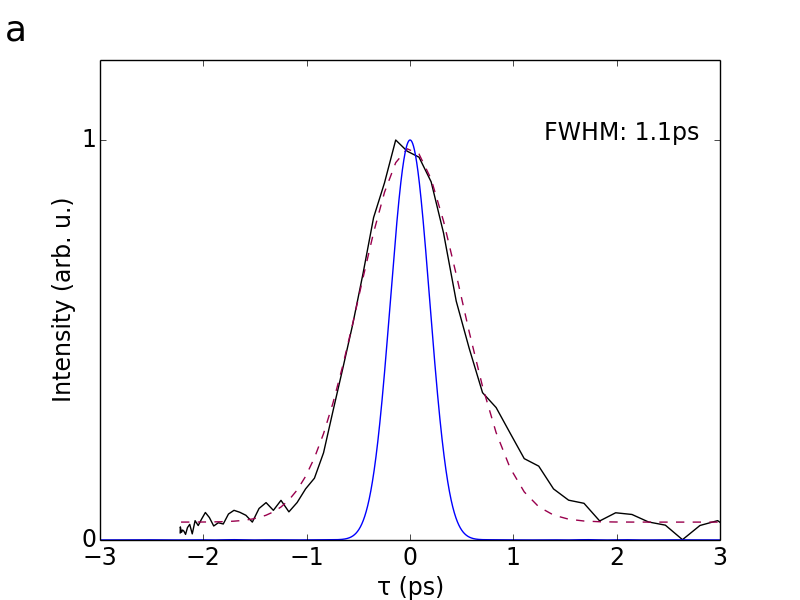}\includegraphics[width=0.45\textwidth]{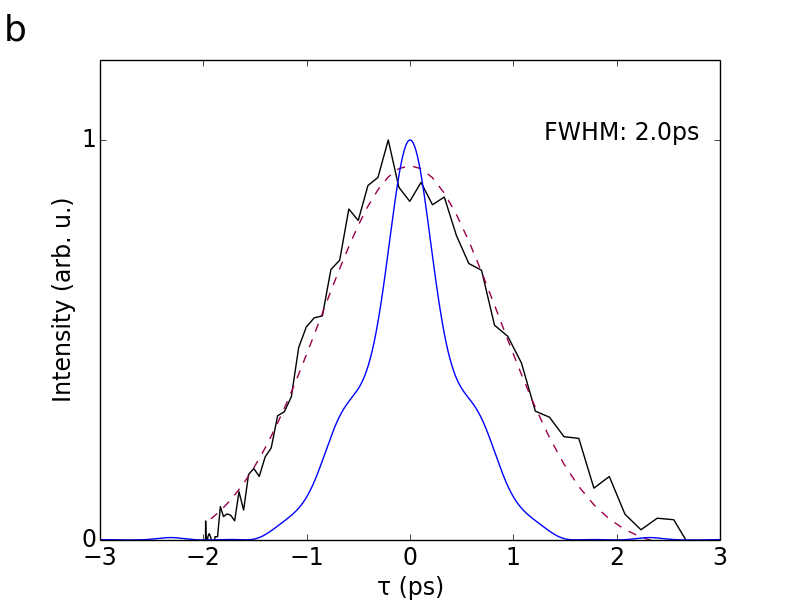}
\caption{Temporal envelopes extracted from the sampling measurement. Solid lines black are data, the dashed lines correspond to Gaussian fits to the data. The solid blue lines correspond to the expected temporal envelope, calculated from cuts through the JSIs, as indicated in Figures~\ref{fig:jsi}a and b.}
\label{fig:timings}
\end{figure}

These states are also characterised with our sampling method.
The background-subtracted temporal envelopes are displayed in figure~\ref{fig:timings} together with the expected temporal envelopes. 
The expected temporal envelopes were extracted from the JSI as the Fourier transform of the conditioned marginal spectrum, i.e. by taking a cut through the JSI along the lines indicated in Figure~\ref{fig:jsi}, and asuming a flat spectral phase (compare Ref. \cite{bre2013}). Only in the case of perfect decorrelation is this cut equal to the complete marginal spectrum.

\begin{equation}
\mathrm{TBP} = \Delta \tau \Delta \nu
\label{eq:tbp}
\end{equation}

The pulse durations extracted from Gaussian fits are \(\Delta\tau\)=1.1\(\pm\)0.2\,ps for the decorrelated state and \(\Delta\tau\)=2.0\(\pm\)0.2\,ps for the correlated one. Now we can calculate the time-bandwidth product for the states as defined by FWHMs in equation \ref{eq:tbp}. The time-bandwidth product is TBP=1.1\(\pm\)0.2 and TBP=1.5\(\pm\)0.2 for the decorrelated and correlated state, respectively. From the measured marginals and the calculated theoretical temporal envelopes we expect TBP=0.57 and TBP=1.1 for the decorrelated and correlated state, respectively.

\FloatBarrier
\section*{Discussion}

While we measure a round joint spectral intensity, which by itself is a weak indication for pure photons, the corresponding TBPs are higher than the Fourier limit. From the increased TBP it is clear that the spectral phase is not flat and there are chirps on the photons. Such chirps could be introduced both after the generation process, or before as a chirp on the PDC pump \cite{sanchez-lozano_relationship_2012}. Both do not show up in the JSI. While a separable chirp introduced after the generation leaves the photon's purity unchanged, a pump chirp has an influence on the multimodeness and of the single photon created \cite{bre2013}. Measuring the temporal envelope together with the spectrum provides additional information the purity over merely taking a JSI.

As the TBP alone does not contain information about phase and pulseshape, and especially not about the source of chirps, we cannot directly calculate back from it.  If the TBP is higher than the Fourier limit, it is not possible to reconstruct if there are non-linear spectral phases present or how multimode the state has gotten. To get the exact purity one would need to calculate a ,,conditioned TBP`` as shown in Ref \cite{bre2013}: Is is calculated from conditioned bandwidths, i.e. the temporal duration of the signal when the arrival time of the idler is fixed, and spectral bandwidths with fixed idler frequency. To do so one needs either both joint spectral and joint temporal intensity, or a means of both spectral and temporal filtering. However, if the state is known, one can calculate both the purity and the TBP. By simulating the PDC source numerically we can carry out this calculation. By attributing any occurring non-linear spectral phases to the PDC pump we obtain a lower bound on the purity.

\begin{equation}
P = tr\left( \rho^2 \right) = \frac{1}{\mathrm{K}}
\label{eq:K}
\end{equation}

On the other hand, by performing a Schmidt decomposition on the measured joint spectral intensities and assuming a flat spectral phase, an upper bound for the purity (see equation \ref{eq:K}) is extracted. For the decorrelated state the Schmidt number K is 1.08 and for the correlated one it is 2.10, yielding purities of 0.93 and 0.48 respectively.

%
%
%

\begin{figure}
\centering
\includegraphics[width=0.9\textwidth]{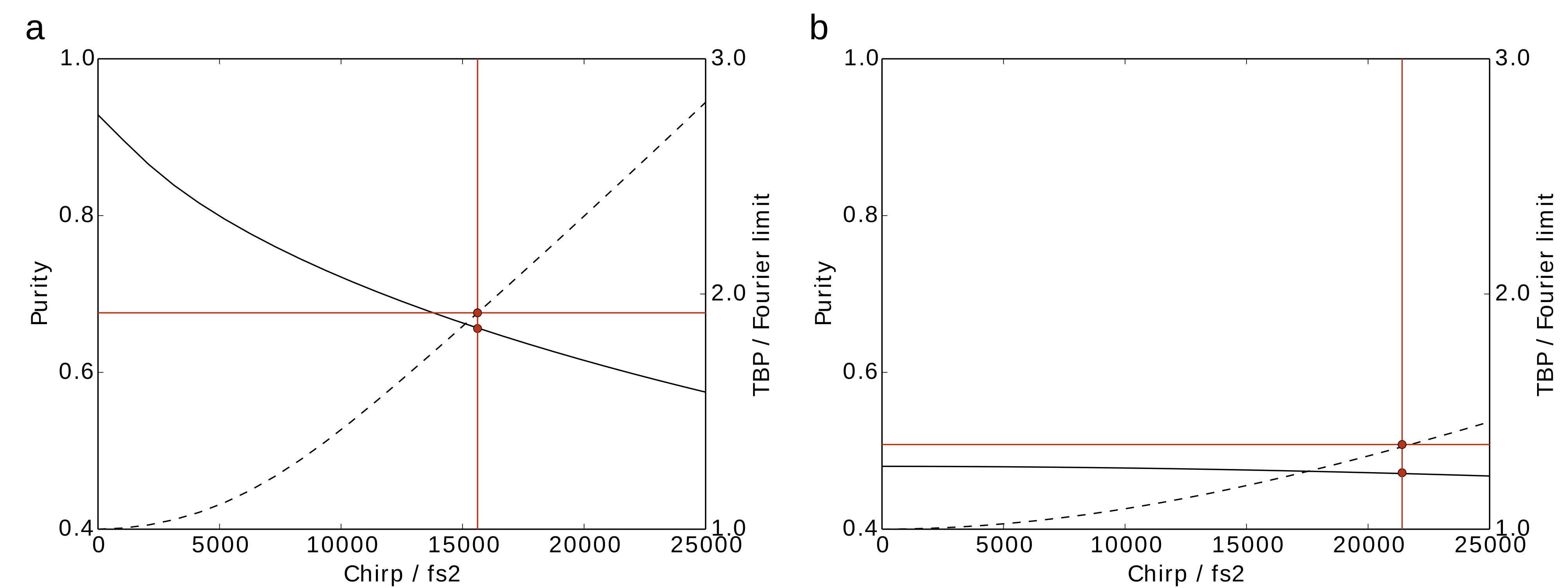}
\caption{Simulated data of a PDC source with the same properties the one employed in this work. The solid line shows purity over pump chirp. The dashed line shown how far the produced state is above the Fourier limit. The dots indicate the point corresponding to the measured state: Panel (a) shows the simulation for the decorrelated state with \(\Delta\nu_{PM} / \Delta\nu_{pump}=1\), panel (b) shows the correlated case with \(\Delta\nu_{PM} / \Delta\nu_{pump}=3.25\)}
\label{fig:theo}
\end{figure}

From the measured JSIs we can infer the bandwidths of phasematching and pump. By simulating the source with the measured parameters numerically we can calculate TBP and purity in dependence of pump chirp. The result is depicted in Figure~\ref{fig:theo} for both the decorrelated state and correlated state. In the experiment we measure a TBP that is 1.92 times the Fourier limit for the decorrelated and 1.36 times higher for the correlated case. From that we can infer pump chirp and purity from the curves plotted. The corresponding pump chirp for the two cases are 15616\,fs\(^2\) and 21400\,fs\(^2\), respectively, where the chirp parameter \(C\) is defined via the quadratic phase term as \(\mathrm{exp} (i \omega^2 C)\). These group delay dispersion values correspond to 0.81\,m or 1.11\,m of fused silica glass, respectively. The purity of the states are 0.656 and 0.472. Interestingly the purity is almost not reduced by the effect of the pump chirp in the correlated case, the decorrelated state is affected much more drastically. These numbers should only be interpreted as an estimate: The numerical modeling needs to be very accurate, and any deviation in term of correlations in the JSI would throw these numbers off. For a precise measurement without the need of further numerical modeling, one would also need a conditioned marginal temporal intensity, or a JTI measurement, as pointed out in Ref. \cite{bre2013}. This would mean to duplicate the entire QPG setup for the second PDC photon. The fact that we obtain a different pump chirp for the two states is a indication that there are either higher order non-linear spectral phase contributions, or the increase of the measured TBP is partially due to separable chirps introduces after the PDC process. Therefore we conclude that the actual purity is between the  upper and lower bounds we ascertained. It is noteworthy that there is a relatively simple way of optimizing a PDC source for maximum singlemodeness. By  measuring the unheralded second order correlation function g\(^{(2)}\)(0), one gets a exact measure of the multimodeness of the sate containing all contributions from PDC pump chirps \cite{christ_probing_2011}. This optimisation has been employed to build PDC sources for highly pure heralded single photons \cite{har2013}

\FloatBarrier
\section*{Conclusions}

We demonstrated a setup for measuring the temporal intensity distribution of single photons from a PDC source with high efficiency. Such  time-domain sampling using a fast delay and long SFG crystals can, if the setup is duplicated for the second PDC photon, also drastically increase measurement speed of the joint temporal intensity measurement as in Ref. \cite{kuzucu_joint_2008}. While the authors in that work reported measurement times in the range of thirty minutes our setup can achieve measurement times of the order of a second, enabling a ,,live view`` of the single photon temporal intensity profile.
With the combination of joint spectral intensity and temporal intensity measurement we were able to identify single photon chirps and with the help of numerical modeling establish concrete upper and lower bounds for the single photon's spectral purity.
If one were to combine two of the devices, one for each PDC photon, the joint temporal intensity could be measured. This completes the characterisation of the spectral-temporal structure and correlations of photon pair states. With knowledge of joint spectral and temporal intensities the phase information could also be reconstructed.

\section*{Acknowledgement}
The authors thank Bastian Reitemeier, Regina Kruse and John Donohue for genuinely useful contributions. This work was funded by the Deutsche Forschungsgemeinschaft via SFB TRR 142, grant number TRR142/1, and via the Gottfried Wilhelm Leibniz-Preis, grant number SI1115/3-1.

\section*{References}

\bibliography{own,bandwidth,Streak,Timing,Telecom,books,spectral,theory}
\bibliographystyle{unsrt}

\end{document}